\documentclass[10pt,letterpaper]{article}

\usepackage{graphicx}
\usepackage{palatino}
\usepackage{xcolor}
\usepackage{ragged2e}
\usepackage{mdframed}
\usepackage{etoolbox}
\usepackage{sectsty}
\usepackage{amssymb,amsmath}
\usepackage{textcomp,textgreek}
\usepackage[hidelinks]{hyperref}
\usepackage[driver=xetex,top=0.85in,left=0.75in,footskip=0.75in]{geometry} 

\usepackage[nospace]{cite}
\usepackage{numberlabel}
\usepackage{citesupernumber}
\makeatletter
\renewcommand{\@biblabel}[1]{\quad#1.}
\makeatother
\usepackage[aboveskip=1pt,labelfont=bf,labelsep=period,
  justification=raggedright,singlelinecheck=off]{caption}

\allsectionsfont{\Large\bfseries}
\subsectionfont{\bfseries\normalsize}
\paragraphfont{}

\RaggedRight
\usepackage{lastpage,fancyhdr}
\fancyfootoffset[R]{1.75in}
\newcommand{\plabel}{\lowercase}
\newcommand{\fig}[1]{Fig~\ref{fig:#1}}
\newcommand{\figp}[1]{(\fig{#1})}
\newcommand{\panel}[2]{\fig{#1}\plabel{#2}}
\newcommand{\panelp}[2]{(\panel{#1}{#2})}

\makeatletter
\g@addto@macro\@floatboxreset\centering
\makeatother

\newcommand{\placefigure}[3]{
  \begin{figure}[tbp!]
    \newgeometry{right=0.5in}
    \setlength{\textwidth}{7.25in}
    \includegraphics[width=0.9\textwidth]{fig_#1.png}\par\medskip
    \caption{\color{darkgray}{\bf #2.}{\small \\ #3}}
    \label{fig:#1}
    \setlength{\textwidth}{5.25in}
    \restoregeometry
  \end{figure}
}

\mdfdefinestyle{box}{%
  backgroundcolor={blue!8},
  innerrightmargin=0.5in
}
\newcommand{\glossterm}[1]{\noindent\textbf{\emph{#1}}\quad}
\newcounter{box}
\newcommand{\bx}[1]{Box~\ref{box:#1}}

\newcommand{\placebox}[3]{
  \begin{figure*}
  \begin{mdframed}[style=box]
    \setlength{\parindent}{-1in}
    \newgeometry{right=1in}
      \refstepcounter{box}
      {\color{darkgray}{\noindent\bf Box~\thebox. #2.}
        \small \\ \noindent #3}
      \label{box:#1}
    \restoregeometry
  \end{mdframed}
  \end{figure*}
}

\begin{document}

\newcommand{\corrauthorname}{Joseph D. Monaco}
\newcommand{\corrauthoremail}{jmonaco@jhu.edu}
\newcommand{\thetitle}{A brain basis of dynamical intelligence for AI and
computational neuroscience}

\hypersetup{%
  pdfauthor=\corrauthorname,
  pdftitle=\thetitle,
  pdfsubject=Manuscript for submission,
  pdfkeywords=artificial intelligence; computational neuroscience; neuroscience; brain;
    neural computation; learning; oscillations; temporal dynamics; attractor dynamics 
}

\begin{flushleft}
{
{ \LARGE\textbf{ \newline \thetitle } }
\\
\bigskip
Joseph~D.~Monaco\textsuperscript{1$\ast$}, 
Kanaka~Rajan\textsuperscript{2}, and 
Grace~M.~Hwang\textsuperscript{3$\dagger\ast$}
\\
\bigskip
\textbf{1} Department of Biomedical Engineering, Johns Hopkins University (JHU)
School of Medicine, Baltimore, MD, USA;
\\
\textbf{2} Icahn School of Medicine at Mount Sinai, New York, NY, USA;
\\
\textbf{3} JHU/Applied Physics Lab, Laurel, MD, USA; JHU Kavli Neuroscience
Discovery Institute, Baltimore, MD, USA.
\\
\medskip


\textsuperscript{$\dagger$} This material is based on work supported by
(while serving at) the National Science Foundation. Any opinion, findings,
and conclusions or recommendations expressed in this material are those of
the authors and do not necessarily reflect the views of the National Science
Foundation.



\medskip

\textsuperscript{$\ast$} Correspondence to
\href{mailto:\corrauthoremail}{\corrauthoremail} or
\href{mailto:grace.hwang@jhuapl.edu}{grace.hwang@jhuapl.edu}.
}
\end{flushleft}


\section*{Abstract}

The deep neural nets of modern artificial intelligence~(AI) have not achieved
defining features of biological intelligence, including abstraction, causal
learning, and energy-efficiency. While scaling to larger models has delivered
performance improvements for current applications, more brain-like capacities
may demand new theories, models, and methods for designing artificial learning
systems. Here, we argue that this opportunity to reassess insights from the
brain should stimulate cooperation between AI research and theory-driven
computational neuroscience~(CN). To motivate a brain basis of neural
computation, we present a dynamical view of intelligence from which we elaborate
concepts of sparsity in network structure, temporal dynamics, and interactive
learning. In particular, we suggest that temporal dynamics, as expressed through
neural synchrony, nested oscillations, and flexible sequences, provide a rich
computational layer for reading and updating hierarchical models distributed in
long-term memory networks. Moreover, embracing agent-centered paradigms in AI
and CN will accelerate our understanding of the complex dynamics and behaviors
that build useful world models. A convergence of AI/CN theories and objectives
will reveal dynamical principles of intelligence for brains and engineered
learning systems. This article was inspired by our symposium on dynamical
neuroscience and machine learning at the 6th Annual US/NIH BRAIN Initiative
Investigators Meeting.

\section*{Main}
\label{sec:intro}

The functional limitations of the current wave of artificial intelligence~(AI),
based on deploying deep neural nets for perception, language, and reinforcement
learning applications, are coming into focus. A recent avalanche of reviews,
perspectives, podcasts, virtual seminars, and keynotes have collectively
signaled remarkable agreement about brain-like capacities that escape our
understanding: abstraction, generalization, and compositionality; causal
learning and inference; cognitive flexibility for generalized problem-solving;
construction of world models; low sample-complexity and high energy-efficiency.
These conversations have not only coursed through the AI and machine learning
communities, but also the neuroscience, psychology, cognitive science, robotics,
and philosophy of mind communities. As AI may be unlikely to bridge these
acknowledged gaps by continuing to simply scale up models, datasets, and
training compute, the opportunity arises to return to the brain for insight.

\looseness=-1 Neuroscientists have begun to integrate deep neural nets into
their methods\cite{StorKrie19,MathMath20} and to search for representations
predicted by deep learning\cite{CadiHong14,YamiDiCa16}. The critical question
is whether this benefit can be reciprocated: Can our current, albeit
incomplete, knowledge of brain-based intelligence translate to meaningful
algorithmic innovation in AI? A corollary question is equally important:
Could stronger partnerships between neuroscientists and AI researchers
help to resolve obstacles in both fields? In this Perspective, we hope to
stimulate the search for the critical set of neurobiological features that
supports the adaptive intelligence of humans and other animals. Not every
biological detail is relevant to the brain's computational capacities, but the
minimal set of idealized neural features in AI leaves the door open to many
potential brain-based innovations. We argue that the still adolescent field of
\emph{computational neuroscience}~(CN) (see \bx{gloss}) should play a key role
in this search.

Given this context, some discussions below might appear overly theoretical.
We embrace the fact that, as neuroscientists, we cannot yet point to certain
physiological details or write down a set of equations as definitive keys to
neural algorithms of intelligence. Instead, we take a step back to synthesize
ideas from biology, neuroscience, cognitive science, and dynamical systems
to outline a brain basis of neural computation that we think points in the
right direction. Thus, we elaborate, in the next section, a working definition
of intelligence, and then consider its neural contingencies by addressing
network structure, temporal dynamics, and future directions based on interactive
learning. These last three sections emphasize distinct perspectives on sparsity.

\placebox{gloss}{Glossary of terminology}{%
  \glossterm{afferent/efferent} providing input to a given target/carrying
      output from a given source \\
  \glossterm{causal invariance class} a set of system states that, conditioned
      on other members of the set, cause the same effect; to support effective
      causal learning and inference, causal invariance with respect to a discrete or
      binary outcome, such as the activation of a downstream reader or token, must
      be understood as an independent causal capacity, i.e., a mechanism, of the
      class itself~[\citen{ChenLu17}] \\
  \glossterm{cell assembly} a fundamental but evolving concept in
      neuroscience that posits connected sets of neurons with topologically closed
      (`reentrant') loops among their synapses that autonomously sustain the cells'
      activation; crucially, Hebb realized~[\citen{NadeMaur20}] that such activation
      would itself modify the loop and potentially find new pathways to distribute, or
      consolidate, the cell assembly~\panelp{structure}{e} \\
  \glossterm{compositionality} the property of symbolic systems that the
      meaning of complex expressions (e.g., sentences) is completely determined by
      syntax and the meanings of simple parts \\
  \glossterm{computational neuroscience} the theoretical investigation of
      computational models of brain function \\
  \glossterm{computationalism} the classical cognitive science theory that
      minds and brains are information processing systems and that cognition should be
      understood as computation \\
  \glossterm{connectome} a detailed network connectivity map of an
      individual brain \\
  \glossterm{dynamical systems} a mathematical approach to the long-term
      behavior of complex adaptive systems as ensembles of particles whose states obey
      differential equations over time \\
  \glossterm{ergodicity} the property of a dynamical system that, from any
      initial state, it will visit all reachable states over the long term, including
      a return to the initial state; thus, ergodic dynamics are not reducible or
      decomposable \\
  \glossterm{ethological relevance} the degree to which a situation or
      experiment aligns with an animal's natural behavior in ecologically appropriate
      environments \\
  \glossterm{hierarchy} a coherent organization of transitive superiority
      relations (`above', `below', or `equal') among elements, typically represented
      as a tree \\
  \glossterm{meaning} internally constructed information with adaptive or
      ethological relevance to an agent or organism \\
  \glossterm{minimal models} an approach to model design that relies on
      theory, intuition, and explanatory power to maximally abstract, idealize, and
      distill complicated systems into a set of essential functions \\
  \glossterm{network of networks} large-scale computational models
      built from complex recurrent neural networks wherein distinct submodules
      simultaneously represent different brain regions \\
  \glossterm{neural inspiration} a vacuous notion that allows nearly any
      system or model to be described as brain-like (cf. `biologically plausible') \\
  \glossterm{physical layer} the material organization of a computer that
      establishes the lowest level of information processing; mechanisms within the
      layer may abstract its raw states into the codes and parameters of computation \\
  \glossterm{preconfiguration} the relative multi-scale stability of
      connectivity patterns in mature brains \\
  \glossterm{pseudohierarchy} our relaxed conception of hierarchy based on
      arrangements of specialists and generalists that allows some degree of level or
      modularity violations~\panelp{structure}{b} \\
  \glossterm{quasiattractor} a local energy minimum that shapes weakly
      convergent flows but nonetheless provides access to divergent flows and
      chaotic states \\
  \glossterm{readers} downstream targets, \emph{viz.} neurons, cell
      assemblies, tokens, or networks, that respond to configurations of states among
      their inputs~[\citen{Buzs10}]; i.e., they read out brain states and contribute
      to expressive transformations of internal sequences \\
  \glossterm{rich club} a set of strongly interconnected generalists that acts
      as a global information hub \\
  \glossterm{small world} the property that average path lengths tend to grow
      only logarithmically with network size \\
  \glossterm{spandrel} an evolutionary trait that is contingent upon other
      adaptive traits but not necessarily adaptive itself \\
  \glossterm{specialists/generalists} neurons or networks with
      high/low prevalence and low/high degrees of convergent (afferent) input; they
      constitute lower/higher `levels' of the brain's pseudohierarchy \\
  \glossterm{systematicity} the combinatoric capacity of a compositional
      system for producing complex expressions \\
  \glossterm{tokens} a mechanistic concept (elaborated below) of a
      computational unit for the brain, grounded in the physical layer of synaptic
      memory loops, that bridges low-level dynamics to functional states \\
}

\section*{A dynamical/behavioral view of intelligence}
\label{sec:intelligence}

While it is difficult to answer ``What is intelligence?'', it is almost
as useful to answer ``What is intelligence \emph{for}?'': Intelligence is
\emph{for} adaptive behavior. Otherwise, an organism would have been better off
(as in the neuromythology surrounding the sea squirt) ingesting its brain and
attaching itself to a rock. A corresponding yardstick for intelligence would be
the degree to which an organism or agent controls its environment in service of
continued survival\cite{Fris13, ParrFris18}. Indeed, extending this assessment
to novel or unpredicted situations, along ecological dimensions, should
correlate with generalized problem-solving capacity\cite{Chol19}.

\looseness=-1 This not-unusual definition of intelligence puts AI (based on
disembodied and nonagentic neural nets trained on datasets lacking spatial,
temporal, epistemic, mnemonic, social, and/or environmental context) at a
disadvantage for purposes beyond hypercompetent regression and classification.
Behavior is variable and complex, but it is also hierarchically organized
through time in all animals, with humans exhibiting perhaps the deepest such
hierarchies. Conceptual knowledge is similarly hierarchical and demanding
of flexibility, reconfigurability, and combinatoric expressiveness (cf.~the
\emph{compositionality} and \emph{systematicity} of language). High-level
cognition is ordered, temporal, and dynamical in that what came before
conditions the \emph{meaning} of what comes after, with lifelong horizons in
both directions.

But where is the computational layer? Network \emph{preconfiguration} and its
metabolic advantages preclude basing this dynamism on first-order mechanisms
of structural plasticity. For instance, conceptual learning would require
regular rebalancing of global connectivity distributions because the linear
`training curriculum' of experience would at most be capable of incrementally
appending leaves to the concept tree during the first (online) stage of standard
two-stage memory models\cite{NadeMosc97,NadeMaur20}. This rebalancing would
become increasingly necessary, due to finer-grained categories, and increasingly
expensive, due to deeper branching. Thus, a different kind of substrate must
efficiently maintain, update, and operate on the hierarchical models in
long-term memory. We focus on temporal and attractor dynamics as the axes of
this computational layer.

\section*{Sparse, skewed structures for flexible abstraction and generalization}
\label{sec:structure}

Typically $<$1–2\% of possible unit-wise connections exist within the
cortico-limbic circuits of the hippocampus and neocortex. The impressive
combinatorics inherent in this level of sparsity\cite{TrevRoll91} give rise
to the intuitive, but perhaps wishful, notion that discovering the underlying
motifs, generating functions, or \emph{connectomes} of synaptic connectivity
will unlock the brain's neural coding secrets. Without such sparsity, dense
connectivity \panelp{structure}{a} either reliably relaxes into pattern
completion for recurrent models \emph{viz.}~Hopfield nets, or universal function
approximation for feedforward models \emph{viz.}~multi-layer perceptrons and
deep learning. Brains appear to do both, but also much more\cite{Sejn20,
BaraKrak21}. Density, as in typical artificial neural nets (ANNs), collapses the
space of possible network configurations to that of size and layer architecture.
Having far fewer degrees-of-freedom greatly restricts structural, and thus
functional, diversity. As brains evolved, such restricted variation would have
shunted the phylogenetic discovery of the inductive biases\cite{SinzPitk19}
that now presumably undergird brain function. If so, structural sparsity is an
ancient precondition for biological intelligence.

\placefigure{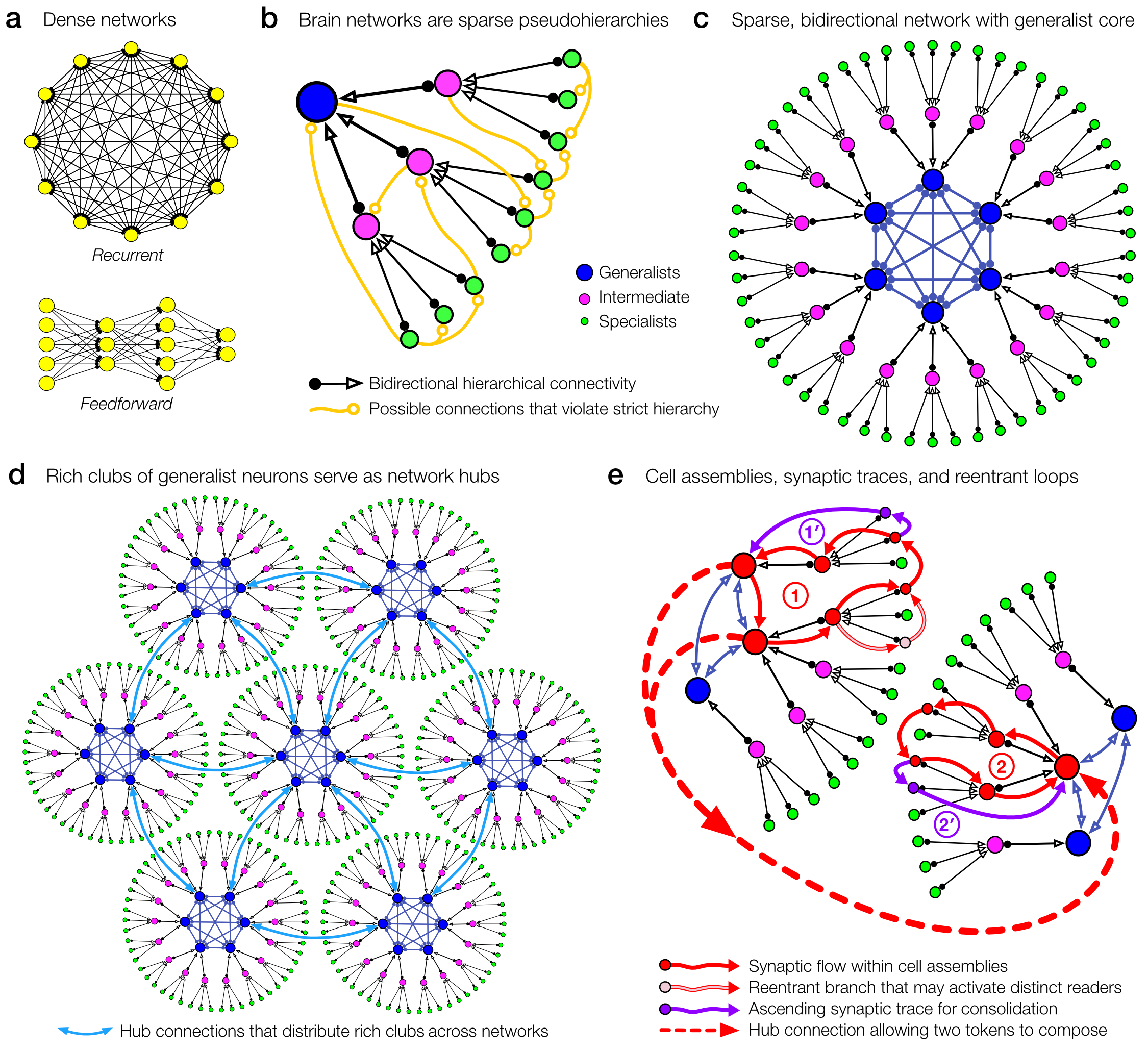}
{Brain networks are sparse, pseudohierarchical, and distributed with log-skewed
connectivity}
{The multi-scale neural structures of the brain constitute its physical layer
of computation. Typical neural net models acquire sparsity via rectified unit
activations, but the brain may take advantage of its sparse pseudohierarchical
structures to amplify the computational capacities of neurons and networks.
\textbf{a}, Dense recurrent and feedforward networks as in Hopfield nets and
deep hierarchical neural nets, respectively.
\textbf{b}, Hierarchies are powerful data structures, but they strictly require
binary and unambiguous superiority relations. Neurons and networks inhabit a
continuum of input connectivity from \emph{specialists} to \emph{generalists}.
Sparse, multi-scale arrangements of elements from this skewed distribution
will form pseudohierarchical structures with lateral and ascending/descending
violations (e.g., level skips).
\textbf{c}, For ease of visualization, we show a balanced hierarchy with a core
subset of strongly interconnected generalist neurons \emph{viz.} a rich club.
Hierarchy violations like those in \textbf{b} are implied.
\textbf{d}, The cortico-limbic system comprises networks linked by sparse,
long-range hub connections that make a small world from rich clubs.
\textbf{e}, The flow of cortical computation emanates from self-sustaining
activity within the reentrant synaptic loops of interconnected cell assemblies
(red circles, 1 and 2). Loops may branch into subloops or aggregate new traces
(purple circles, 1$^\prime$ and 2$^\prime$) that entail different effects on
downstream targets, thus instantiating distinct \emph{tokens} of neural
computation (see below).}

\subsection*{Obstacles for network models with sparse connectivity}

The recognized importance of anatomical/structural sparsity has taken distinct
forms in CN/neuroscience and AI. In neuroscience, a current approach posits that
commiditizing connectomes (in the tradition of the genetic sequencing project)
will unlock crucial new technologies and potential therapies\cite{FornZale15}.
Recent advances in AI, including network distillation\cite{BurdEdwa18a},
lottery tickets\cite{FranCarb19}, and synaptic flow\cite{TanaKuni20},
have wielded structural sparsity to reduce model complexity in light of
Sutton's cautionary note\cite{Sutt19} and concerns about training extremely
large models, such as `double descent'\cite{NakkKapl19} and environmental
sustainability\cite{ThomGree20}. These reactions reflect the fields’
respective mainstream interests: neuroscience wants neurotechnology capital to
keep funding big labs and consortia\cite{LitvAdam19, HsuFang20, Wool20}; AI
wants more efficient training to quickly deploy updated models for its trending
applications\cite{BrowMann20, SchiSchu21}. 

First, generalizing from frozen, or static, connectivity patterns is complicated
by the `synaptic weight', one of the two main parameters tuned when training an
ANN (along with unit bias). The resulting weight matrices determine the gains of
unit-wise directed connections in a network. However, the gains of biological
synapses as measured from imaging or electrophysiology are in constant flux
due to, e.g., homeostatic and neuromodulatory brain states that complement
experience-dependent learning in nontrivial ways\cite{MongRump17}. At best,
synaptic volatility obscures the functional relevance of any particular weight
matrix, or learning rule, over meaningful periods of time. In CN modeling,
an inherent trade-off between the level of data-driven biological detail
and empirical capacities for inference and generalization\cite{LeveAlva20}
forces strong assumptions to be made about behavioral and functional
states. These factors preclude highly efficient ANN idealizations like
rectified-linear unit activations\cite{GlorBord11} and backpropagation of error
(`backprop')\cite{AhmaSche19}.

Second, whereas synaptic weights fluctuate, the overall pattern of connectivity
(in mature, developed brains) is less dynamic. This might appear to boost
the value of frozen connectomes, but evidence for numerous conserved,
functional brain states across individuals and species through waking and
sleep\cite{McCoNest20, JacoStei20} implies that any particular connectome
(within healthy interindividual variability) underdetermines its associated
functional states. Why should this be the case? If metabolic efficiency were
an evolutionary driver of structural sparsity\cite{AttwLaug01, LevyCalv21},
then the energetics of functional states that continually reorganized axon
collaterals, synaptic boutons, dendritic arbors, etc., would surely be selected
against.

CN network models examine conditions of theory-driven, kernel-based, or
sparse-random connectivity, wherein sparsity is typically around 5-10\%
due to the breakdown of smaller (i.e., computationally feasible) models at
more brain-like sparsity. Methods for scaling up CN models include partial
inference such as isolated cell-type pre-tuning\cite{MarkMull15}, multi-region
\emph{network of networks} models\cite{PeriRaja20}, and formal (e.g., mean
field or master equation) approaches to mesoscopic dynamics\cite{LeenFrie12,
ZerlChem18}. Further, CN models typically study learning rules based on Hebbian
association (or similar) and the types of local plasticity mechanisms that have
been the focus of experiments. Thus, global update rules like backprop have only
recently renewed theoretical attention in neuroscience\cite{LillSant20}.

\subsection*{It's a log-log world (after all)}

Which properties of cortical connectivity form the computational layer of
dynamical intelligence? We highlight two points about connection structure.
First, in a sparse cortex, one effect of surprisingly low average path
lengths\cite{Spor10} is global inter-regional accessibility. This \emph{small
world} network property\cite{WattStro98} can emerge developmentally, via
activity-dependent pruning, from simple rules in dynamically synchronous
populations\cite{KwokJuri06, JarmTren14}. The resulting logarithmic scaling
of \emph{efferent} connections forms coherent local neighborhoods with
efficient access to any other neighborhood in the cortical sheet. Second, in a
hierarchical cortex, the connectivity graph, between neurons or minicolumns,
need not be strictly isomorphic to a tree in the computer science sense. In
fact, a cortical B-tree would be unacceptably fragile toward the root node due
to poor distribution of connections.

Indeed, general conceptual knowledge and remote memories are more robustly
accessible than their specific and recent counterparts, likely from having
aggregated multiple traces, grounded in the medial temporal lobe and
hippocampal-entorhinal cortical complex\cite{NadeMosc97, Bars99, HaisGore01,
MokLove21}. Consequently, while we refer to the cortico-limbic long-term memory
graph as a \emph{hierarchy}, systems consolidation implies that the graph is
more like a distributed multi-scale arrangement of neurons and networks along a
continuum of \emph{afferent} input convergence, i.e., from \emph{specialists}
to \emph{generalists} \panelp{structure}{b}. Indeed, studies of mature
preconfigured memory networks revealed, again, log-skewed distributions with
long tails of more excitable generalist neurons\cite{BuzsMizu14, Buzs19,
McKeHusz21}. The generalist neurons and networks can organize, despite their
smaller numbers, into highly stable \emph{rich clubs}\cite{HeuvSpor11,
BiniStri15, DannMich16, ReesWhee16} that serve to robustly distribute the higher
levels of the hierarchy \panelp{structure}{c}. Thus, small~world output (i.e.,
log-skewed efferent access) and rich~club input (i.e., log-skewed afferent
tuning) may constitute a slide-rule-like \emph{physical layer} of computation
\panelp{structure}{d} for flexible abstraction and generalization.

\section*{Complex temporal dynamics for computational sequences of sparse states}
\label{sec:temporal}

In contrast to the timing agnosticism of classical \emph{computationalism},
the dynamical/behavioral view of intelligence prioritizes timing (vs.~order):
The right behavior at the wrong time is equally deadly as the wrong behavior,
because it is nonetheless coupled to the rest of the world. That the phylogeny
of biological intelligence is a story of interacting with the world emphasizes
cognition as an internal physical process that unfolds through time to manage
this inextricable coupling to external forces. The analysis of mechanistic
couplings over time is the domain of \emph{dynamical systems} theory. Thus, a
dynamical systems perspective has emerged\cite{SkarFree87, Geld95, Geld98a,
Free03, Izhi07, Milk13, KozmFree15, RabiSimm15, JonaKord17, Milk18, RabiVaro18,
MastOsto18, DannAlex19, VyasGolu20} within cognitive science, philosophy of
mind, and neuroscience, wherein ``[c]ognition is then seen as the simultaneous,
mutually influencing unfolding of complex temporal structures''\cite{Geld98a}.

Respecting the transparency of temporal variation unlocks a crucial dimension
along which to organize and interrelate neural events. Continuous-time networks
of dynamical spiking neuron models robustly demonstrate self-organized
synchronous groupings that expand functional capacities\cite{Izhi06, KilpErme11}
\panelp{timing}{a} and provide a stronger causal basis for empirical
explanation\cite{Bret12, Bret15a} compared to discrete, rate-based ANNs and
similar \emph{minimal models} in CN. The `rate~vs. time' argument about neural
coding goes back at least 50 years\cite{PerkBull68} and has mostly revealed
new ways for theorists to talk past each other. Thus, the dynamical systems
view emphasizes spike timing (vs.~firing rates) because temporal relationships
(1)~allow causal mechanisms to continuously unfold in sequence\cite{Geld98a} and
(2)~avoid the observer bias inherent in calculating time-binned average firing
rates\cite{Bret15a, Bret19}.

\placefigure{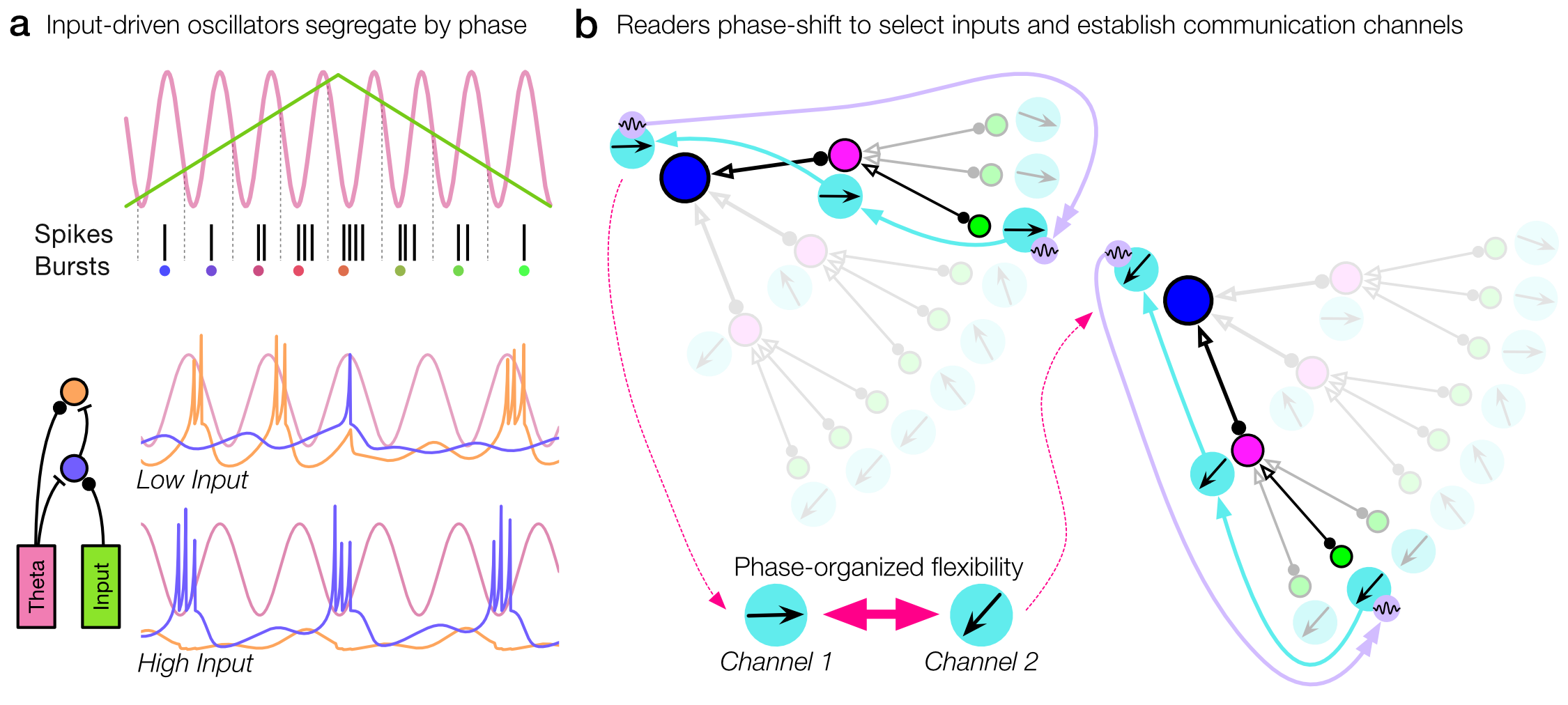}
{Phase synchronization and nested oscillations can sequence, segregate, and
communicate}
{Robust neural mechanisms for transforming inputs into timing signals aligned
to ongoing oscillations (i.e., phase codes) may organize neural activity into
computationally ordered sequences and inter-regional communication channels.
\textbf{a}, (Top) The spike timing of a model neuron reveals a phase-rate
code for a slowly ramping input (green). This phase dependence arises from an
oscillatory input (inhibitory, pink) such as from a theta-rhythmic interneuron.
(Bottom) A simple extension of this model to a circuit with two bursting cell
types (purple vs. orange, left) demonstrates that input level controls the
oscillatory phase of bursts emitted by the circuit (133-ms theta cycles,
right). Adapted from Monaco et al.~(2019)~[\citen{MonaDe-G19}] as permitted by
the CC-BY 4.0 International License (creativecommons.org/licenses/by/4.0/).
\textbf{b}, Nested oscillations consist of a fast oscillator (FO) whose
amplitude is modulated by the phase of a slow oscillation (SO). The broader
spatial coherence of the SO means that it may influence more remote levels of
the cortical pseudohierarchy \panelp{structure}{b}. For example, a communication
request may be initiated by a specialist (green) that successfully resets the
SO phase-angle (cyan circles with arrow) of a generalist (blue), which then
customizes a FO sequence (purple double-arrow) for the input-receptive SO phase
of the specialist. Readers, like this generalist, may use phase-amplitude
coupling to flexibly and selectively communicate with their inputs (e.g.,
\emph{Channel~1} vs. \emph{Channel~2}). Thus, temporal dynamics within an
oscillatory hierarchy may support crucial neurocomputational capacities.
}

\subsection*{Oscillations as reentrant flows on recurrent networks}

If interaction is the net effect of intelligence and behavior is deeply
hierarchical, then we might expect the neural mechanisms of intelligence to
vary over a hierarchy that is functionally isomorphic to that of behavior. By
construing this variation as temporal, oscillations and neural synchrony become
candidates for this isomorphism, particularly on the basis of oscillatory
nesting between timescales~\panelp{timing}{b}. To illustrate the \emph{why} and
\emph{how} of functional oscillations, we sketch several findings:
\begin{itemize}
  \item Observed frequency bands span from circadian (1~day) and ultradian
    (90~min) periods through infraslow (1/10 Hz) and slow (1~Hz) cycles up to
    theta \mbox{(4--8~Hz)}, alpha \mbox{(8--12~Hz)}, beta \mbox{(12--30~Hz)},
    and gamma \mbox{(30--100~Hz)} oscillations, and transient ripples
    \mbox{(100--200~Hz)}\cite{BuzsLogo13};
  \item The frequency ratios between successive bands are approximately
    ($\sim\!e$) constant\cite{PentBuzs03}; i.e., they follow logarithmic
    intervals, minimizing overlap and harmonic interference;
  \item Slower oscillations maintain spatial coherence over larger regions, 
    and no mammalian oscillation has been found that was not nested in the
    cycles of a slower oscillation, e.g., via the phase-amplitude form of
    cross-frequency coupling (CFC)\cite{CanoEdwa06, AruAru15, HyafGira15,
    Buzs19}, thus forming a spatiotemporal hierarchy;
  \item The foregoing properties, including approximate frequencies, are
    evolutionarily conserved across mammals, scaling from the brains of shrews to
    baleen whales over three orders-of-magnitude\cite{BuzsLogo13, Buzs19};
  \item Detailed CN models of the locally recurrent excitatory/inhibitory
    (E/I) networks prevalent in cortico-limbic areas demonstrate robust
    emergence of rhythmic synchrony based on relative E/I gains and synaptic
    time-constants\cite{WangBuzs96, WhitTrau00, GeisBrun05, BuzsWang12, VerdBodn12};
  \item Input drive may balance E/I currents to produce \emph{desynchronized}
    (e.g., low-amplitude gamma) activity\cite{HennVoge14, ZerlDest17, AhmaMill19}
    in networks that otherwise relax into \emph{synchronized} (e.g., high-amplitude
    alpha) states\cite{PfurStan96, PariHans18}.
\end{itemize}
In AI, attempts to incorporate timing have reductively mapped stochasticity
to dropout\cite{KrizSuts12}, continuous dynamics to spikes or binary
activation\cite{LeeDelb16, TavaMaid19, KherMirs20}, and nonlinear recurrence to
memory gates\cite{HochSchm97, ChunGulc14}. In the brain, however: variability
should not be confused with noise; spikes can be understood as autonomous
oscillations that can be nested within bursts or ripples\cite{Buzs15,
KayFran19}, suggesting they are the root of the oscillatory hierarchy;
and biological recurrence foments the chaos of total history-dependence,
presciently described in 1919 by the French zoologist Yves Delage (as quoted by
Buzs\'aki [\citen{Buzs19}, p.~85]), ``the neuron's vibratory mode as a result
of its coaction with others leaves a trace that is more or less permanent in the
vibratory mode resulting from its hereditary structures and from the effects of
its previous coactions.''

What is the function of hierarchically nested oscillations? Dynamics unfolding
through time are about sequences and sequences are about computationally ordered
trajectories of states. That is, nested oscillations may \emph{read out chunks}
of the long-term hierarchical models described above. Brain oscillations
are weakly chaotic, meaning that the slower oscillation of a nested pair
can quickly ($<$1 cycle) entrain a remote brain area into a sender/receiver
channel defined by the direction of nonzero phase lags\cite{Colg15, BastVezo15,
HelfKnig16, Eich17, Buzs19}. Over large networks, directional gradients in
coupling frequency\cite{LundBast20} might coordinate macroscale dynamical
flows as traveling or rotating waves\cite{PateFuji12, ZhanJaco15, MullPian16,
ZhanWatr18, HernCoop20}. A recent intervention study of transcranial stimulation
in humans showed that prefrontal cognitive functions were distinctly modulated
by CFC-like stimulation in separate bands\cite{RiddMcFe21}. Thus, while
large waves (and nonspecific modulation) broadly promote synchrony, temporal
phase-organization is what allows readers to flexibly select from among their
inputs\cite{MonaDe-G19} \panelp{timing}{b}.

\subsection*{Syntactic causal tokens as the unit of cortical computation}

The above findings suggest a candidate function for dynamical sparsity:
Temporal dynamics sparsely activate discrete, distributed \emph{tokens}
from the structural world model, and then recursively, compositionally,
and systematically sequence those tokens into higher-order cognitive
processes (cf.~theories of neural syntax\cite{Buzs10, HyafGira15}, linguistic
construction schema\cite{BickSzat09, SteeSzat18}, and conceptual cognitive
maps\cite{ThevFern19, RikhGoth20, PeerBrun21, MortPres21}). To motivate this
concept of tokens, we note that a dynamical unit of computation should be
discrete (i.e., bounded in state space), syntactic (i.e., intrinsically formal),
and mechanistic (i.e., grounded in causal interactions of its substrate).
Tokens encapsulate active computational states, in contrast to latent memory
states considered as reentrant loops in the synaptic pathways of cortico-limbic
networks \panelp{structure}{e} \emph{viz.} the Lashley-Hebb \emph{cell
assembly}\cite{NadeMaur20}.

We define tokens as classes of syntactic neural states that (1)~transiently
self-sustain activation and (2)~competitively suppress accessible successor
tokens. We posit that token discretization arises from the activation of
structurally segregated latent memory states into dynamically integrated
\emph{causal invariance classes} with respect to downstream effects
\panelp{structure}{e}. For a given latent memory state, a token is the
class of active states that \emph{ergodically} self-reinforces the
reentrant flow through its synaptic pathways, conditioned on its targets,
or \emph{readers}\cite{Buzs10}. Thus, tokens are discrete functional states
guided by multi-scale \emph{quasiattractors} of the brain's complex energy
landscape\cite{Miln85, TsudUmem03, CossAron03, LundRehn06, KnieZhan12,
DemiHeus17, KrotHopf18, MonaDe-G19, McKeHusz21, GardHerm21, BaraKrak21,
SiegJia21}.

Why are neural tokens necessary? First, the causal invariance constraint
serves to mechanistically ground the computational layer\cite{Milk13, Milk18}
and provides essential scaffolding for adaptive causal learning\cite{ChenLu17}.
Second, to embed the Lashley-Hebb cell assembly\cite{NadeMaur20} into the
oscillatory hierarchy described above, we needed to recast its assumption
of binding via simultaneous firing\cite{Miln57}. Given temporal dynamics
beyond simultaneity, instantaneous active states are no longer identified with
their latent memory states. Tokens span this ontological gap by mapping the
periattractor transits of dynamical microstates to causal units of computation.

\subsection*{Computing with flexible, composable sequences of quasiattractors}

\looseness=-1 If neural tokens are indeed the functional unit of cortical
computation and cognition, then their attractor-driven stability must
be continually broken. A recent contraction-theoretic analysis showed
that dynamic stabilization of sparse recurrent networks may require
nonassociative (anti-Hebbian or inhibitory) plasticity\cite{KozaLund20};
however, CN models have shown that short-term plasticity, including synaptic
depression\cite{BaraTsod07, RomaTsod14} and neuronal adaptation\cite{DecoRoll05,
AkraRuss12, AzizWisk13, StelUrda20}, can robustly destabilize active states
and facilitate sequence production, even without the asymmetric connectivity
relied upon by earlier models of sequence learning. Yet, how is the next
token selected for activation? Flexible, expressive sequence generators must
have access to novel, divergent paths, but the neurocomputational mechanisms
of this access remain unclear. Several theoretical frameworks and models
provide useful insights, including distributed inference\cite{Brai86}, local
context\cite{Levy96, MonaLevy03}, modular latching\cite{AkraRuss12, SongYao14},
metastability\cite{TognKels14}, winnerless competition\cite{RabiVaro18},
and chaotic itinerancy\cite{Tsud15, Mill16}, wherein the orbits of neural
tokens might follow quasiattractors that fluctuate within nested oscillatory
cycles. Nonetheless, these bottom-up sequences will necessarily be guided
and sculpted by cortical control flow, perhaps implemented by minicolumn
circuits\cite{GeorLaza20a}, to support inference, composition, and other
cognitive processes. Thus, the waking cyclical production of internal token
sequences provides the ``bicycle for the mind'' for the brain (apologies to
S.~Jobs). Riding this bicycle will require new theories and models of joint
temporal-attractor dynamics in biological and artificial systems. For instance,
a recent model from two of the authors demonstrated self-organized swarm control
with phase-coupling and attractor dynamics\cite{MonaHwan20}.

\section*{Future directions: Breaking the learning impasse with sparse,
interactive behavior}
\label{sec:future}

\subsection*{A tale of two fields}

It is not enough to abstract or idealize some function, like
attractor sequences, to declare a biological basis of intelligence.
Miłkowski~(2013)\cite{Milk13} presents a mechanistic account of neural
computation by requiring complete causal descriptions of the neural phenomena
that produce those functions, because those capacities arise from constitutive
mechanisms grounded in brain organization (of which, e.g., the connectome
is only one aspect). In this account, computational functions must ``bottom
out''\cite{Milk18} within mechanistic sublevels to both isolate and limit the
external dependencies of computational states. If this is the case, then our
history of scientific uncertainty about the organizational levels wherein
cortical computations bottom out means that the minimalist \emph{neural
inspiration} for early connectionist ANNs was likely an early-stopping mistake.
Given the state of brain science in the time of Pitts and Rosenblatt, and even
the 1980s, connectionist abstractions necessarily settled at more general, less
explanatory levels.

AI models evolved to substitute many-layered hierarchies for biological
complexity, while protecting the critical path of backprop because of its
``unreasonable effectiveness'' (apologies to R.~Hamming). That strategy is
paid for by the exorbitant training complexity of computational systems that
are essentially open in high dimensions (i.e., blank slates); several recent
models showed that this trade-off can be mitigated with relevant biological
priors\cite{GeorLehr17, SinzPitk19, DalgMill21}. However, the use of technology,
in introducing GPUs to avoid the von~Neumann bottleneck\cite{KrizSuts12},
triggered an entrenchment of hardware and software codesign that at least rhymes
with history\cite{Hook20}. The impressive recent progress in AI applications
has conditioned its models on backprop and a competitive benchmarking culture.
Without relaxing those conditions, the search for qualitatively new models will
remain disincentivized\cite{StanLehm15} and inductive theoretical interpretation
will suffer as \emph{spandrels} become more common, such as the inexplicable
outperformance of the Adam optimizer with default parameters. 

Experimental neuroscience has become similarly trapped. Advancing
neurotechnologies\cite{HsuFang20} have entrenched the blinkered reductionism of
necessary-and-sufficient circuit explanations for experimenter-relative
`behavior'\cite{GomePato14, Gome17}. The linear causality implicit in
this paradigm provides inferences without guarantee of 
\emph{ethological relevance}. Thus, we must admit the circular causality of
dynamical control and action-perception loops that subserve the behavioral
teleology of intelligent animals\cite{KrakGhaz17, Niv20}.

\subsection*{Interactive, agentic learning is global learning}

Animals are agents and, as such, have high-level goals; they behave so as to
reliably achieve those goals\cite{Powe73, Bell14, MusaUrai19}. This agentic
view favors a system of goal-directed behavior based on a simultaneous
coupling of internal simulation (i.e., prediction) and external interaction
(i.e., error correction). Indeed, a predictive coding analysis of canonical
cortical circuits revealed that beta and gamma oscillations may, respectively,
carry such prediction and error signals\cite{BastUsre12}. This oscillatory
division-of-labor is consistent with a ``spectral connectome'' for internal
generative models\cite{Fris08, FrisBast15, DannMich16, ParrFris18, IssaCadi18}
and a recent striking discovery of a beta-band hippocampal mode that emits
reversed sequences\cite{WangFost20}. Corroboration of unifying theories may
require new methods and practices for running experiments in naturalistic
environments that promote authentic interactions, as demonstrated by an
innovative study of rats that happily learned to play hide-and-seek with
human experimenters\cite{ReinSang19}. Open-loop experiments like this can
provide rich complementary datasets to help constrain complex models and
improve the generality of causal inferences in neuroscience. Recent advances
in AI have embraced agentic interaction, rich environments, and especially
play\cite{KosoColl20}, but these feats have depended on rule-based games with
quantified rewards\cite{SilvSchr17, BadiPiot20}. Humans and other animals play
lifelong games involving unpredicted situations and ambiguous or unquantifiable
outcomes. Thus, both AI and CN would benefit from the roboticists' focus on
how agents actively construct meaningful world models and prioritize behaviors
according to self-confidence in beliefs and predictions.

\placefigure{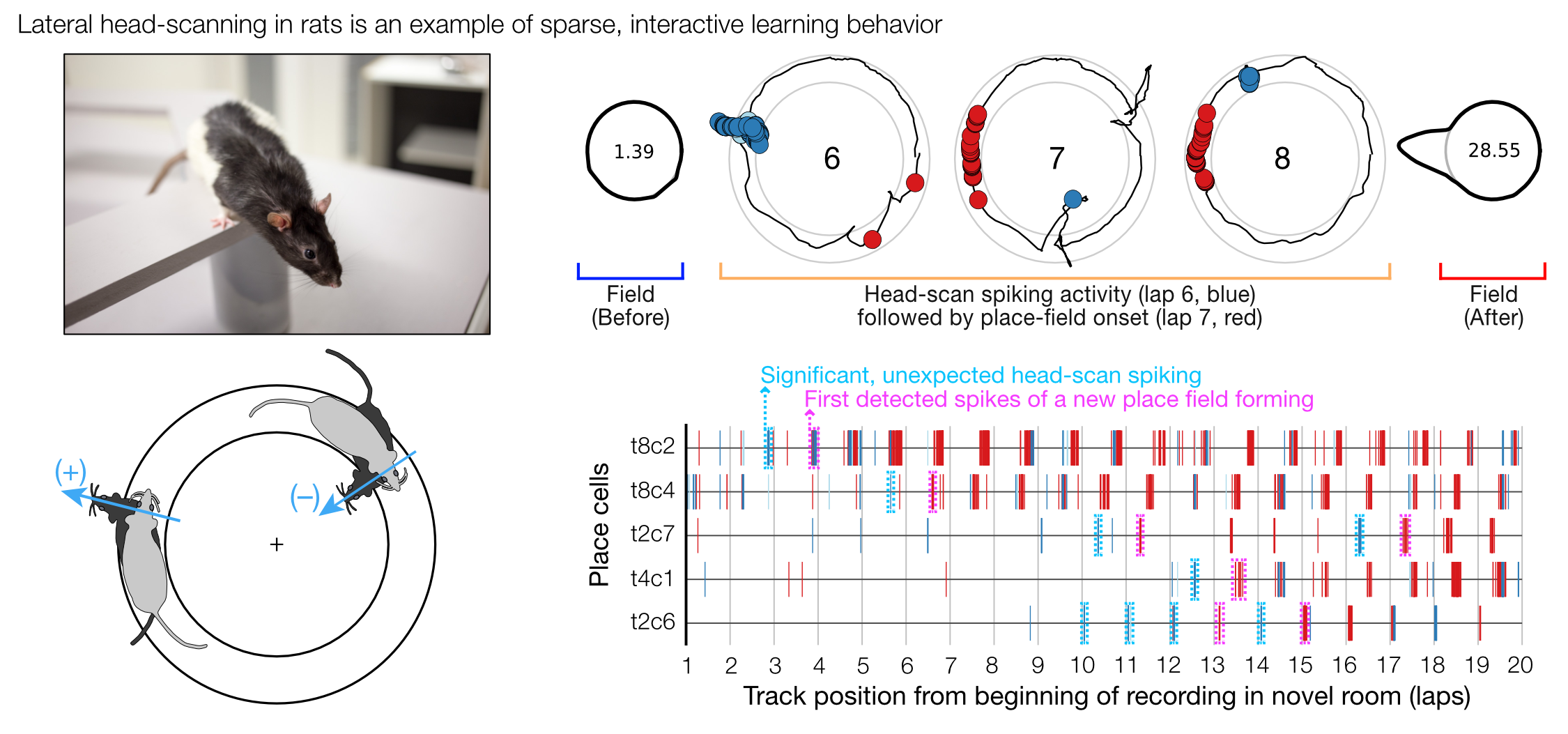}
{Animals are agents that use behavior to learn from discrete interactions}
{We illustrate the punctuated, action-oriented nature of biological
learning by describing a study of lateral head-scanning and hippocampal place
fields in rats (Monaco et al.~(2014)~[\citen{MonaRao14}]). To elucidate
long-term episodic memories, it is critical to understand how cognitive maps in
the hippocampus change over time.
(Photo) Behavioral observations of rats on circular tracks led to the hypothesis
that lateral head-scanning movements during pauses in running may influence
place cells.
(Bottom left) Quantification of pause behaviors allowed isolation of head scans.
(Top right) Place-cell activity was quantified on a lap-to-lap basis to detect
the initial formation or abrupt strengthening of individual place fields.
Statistical analysis demonstrated a highly predictive and specific relationship
between unexpectedly high levels of spiking during a head scan and the formation
or potentiation of a place field at that location on the next lap. Example plots
show scan spikes (blue) and place-field spikes (red) for a scan-potentiation
event on lap 6 (cf. radial firing-rate plots, before vs. after; peak rate in
spikes/s).
(Bottom right) By unrolling 20 laps of spikes from 5 simultaneously recorded
place cells in a novel room, it can be seen that every instance of strong scan
firing (blue dotted boxes) was followed by a new place field (pink dotted boxes)
on the next lap. Thus, over several minutes, these cells were only recruited
into the emerging map in precise conjunction with volitional, attentive actions
of the rat.
}

Animals learn continually, but not continuously. Experience is punctuated
by learning bouts driven by conjunctions of global states including arousal
and attention. These global states emanate from the subcortex, particularly
highly-conserved brainstem structures that modulate awareness and intentional
action\cite{Merk07a, SolmFris18, MunnMull21}. For example, one of the authors
previously showed that rats' hippocampal place-cell maps are abruptly
modified during discrete, attentive head-scanning behaviors\cite{MonaRao14}
\figp{learning}. Recent CN models have shown that this kind of sparse,
action-driven plasticity may facilitate active inference for learning
parsimonious and effective world models\cite{TschSeth20, MartSchu20}. In
contrast, neuroscientific knowledge is most complete for local plasticity
mechanisms at the microscales of transcription, biophysics, and synaptic
ultrastructure. These low-level details are compounded by additional
complexities of cell-type diversity, the role of microglia in pruning
connections, nonlinear integration in dendritic arbors, etc. Whereas AI
profits from the global efficiency of backprop, CN has allowed neurobiological
complexity and a bias for local associative plasticity to obscure theories of
global learning in the brain. Steps toward an agentic modeling paradigm for CN
and (non-reinforcement learning) AI might include global, ethologically relevant
learning signals from agent-equivalents of volitional behaviors, sympathetic
arousal, homeostatic errors, or attentional saliency.

\subsection*{A dynamical future for AI and computational neuroscience}

A recently announced US/NSF Emerging Frontiers in Research and Innovation
program, called BRAID\cite{Braid21}, will be positioned to support innovative
convergent approaches to engineered learning systems with brain-like energy
efficiency based on codesign of neural theories, algorithms, and
hardware, the latter of which has undergone rapid iteration based on innovations
including hardware simulation, event-based sensory inputs, and structured
dendrites\cite{RoyJais19, Mead20, Aimo20, ZenkNeft21, ZenkBoht21}. Promising
neural adaptations of backprop include the relaxation dynamics of equilibrium
propagation\cite{ScelBeng18, MillTsch20a}; compartmentalization of errors in
dendrites, cell populations, or feedback pathways\cite{GuerLill17, BartSant18,
LillSant20}; and gradient descent based on diffusive cell-type-specific
neuropeptide signals\cite{LiuSmit21}. CN research would likewise benefit
from increased adoption of computational practices from AI to promulgate
transparency, reproducibility and replicability, and large-scale model inference
\emph{viz.} automated testing and hyperparameter optimization, architecture
search, regularization, and cross-validation. A mechanistic understanding of
the intertwined concepts of sparsity in network structure, temporal dynamics,
and learning behaviors will help unravel the biological basis of cognitive
computations in humans and other animals. An AI/CN consilience will accelerate
our discovery of shared dynamical principles of intelligence.

\section*{Acknowledgments}

This work was supported by NSF NCS/FO 1835279 (JDM and GMH), NSF NCS/FO 1926800
(KR), BRAIN Initiative NIH NINDS UF1NS111695 (JDM), and NIH NINDS R03NS109923
(JDM). Further support to KR was provided by the Understanding Human Cognition
Scholar award from the McDonnell Foundation. Further support to GMH was provided
by the JHU Kavli Neuroscience Discovery Institute, JHU/APL Innovation and
Collaboration Janney Program, and National Science Foundation (see author
footnote). This article was inspired by open-ended and elucidating discussions
with our BRAIN Initiative symposium panelists, Xaq Pitkow, Brad Pfeiffer, Konrad
Kording, and Nathaniel Daw, all of whom commented on early versions. The authors
thank Patryk Laurent for insightful feedback on the manuscript.

\pagebreak

\label{end_main_document}

\end{document}